\documentclass{article}

\usepackage{amsmath,amssymb}
\usepackage{setspace}
\usepackage{tocloft}
\usepackage{mathtools}
\usepackage{booktabs}
\usepackage{multirow}
\usepackage{bm}
\usepackage{xcolor}

\usepackage{setspace}

\usepackage[margin=2cm]{geometry}
\usepackage{newtxtext}
\usepackage{newtxmath}

\usepackage[absolute]{textpos}

\usepackage{tabstackengine}
\setstacktabbedgap{1.5ex}
\setstackgap{L}{1.2\normalbaselineskip}

\DeclarePairedDelimiter{\norm}{\lVert}{\rVert}
\DeclarePairedDelimiter\floor{\lfloor}{\rfloor}

\cftpagenumbersoff{figure}
\cftpagenumbersoff{table} 

\newcommand{\z}[1]{\textcolor{black}{#1}}
\newcommand{\y}[1]{\textcolor{black}{#1}}

\title{Learning convergence prediction of astrobots in multi-object spectrographs}
\author{Matin Macktoobian, Francesco Basciani, Denis Gillet, and Jean-Paul Kneib%
\date{The authors are with the School of Engineering, Swiss Federal Institute of Technology in Lausanne (EPFL), Lausanne, Switzerland (e-mail: matin.macktoobian@epfl.ch}}%

\begin{document} 
	
	\begin{textblock}{14}(2,0.7)
		\noindent\textbf{\color{red}Published in ``Journal of Astronomical Telescopes, Instruments, and Systems'' \\DOI: 10.1117/1.JATIS.7.1.018003}
	\end{textblock}
	
	\maketitle
	\begin{abstract}
		Astrobot swarms are used to capture astronomical signals to generate the map of the observable universe for the purpose of dark energy studies. The convergence of each swarm in the course of its coordination has to surpass a particular threshold to yield a satisfactory map. The current coordination methods do not always reach desired convergence rates. Moreover, these methods are so complicated that one cannot formally verify their results without resource-demanding simulations. Thus, we use support vector machines to train a model which can predict the convergence of a swarm based on the data of previous coordination of that swarm. Given a fixed parity, i.e., the rotation direction of the outer arm of an astrobot, corresponding to a swarm, our algorithm reaches a better predictive performance compared to the state of the art. Additionally, we revise our algorithm to solve a more generalized convergence prediction problem according to which the parities of astrobots may differ. We present the prediction results of a generalized scenario, associated with a 487-astrobot swarm, which are interestingly efficient and collision-free given the excessive complexity of this scenario compared to the constrained one.
	\end{abstract}
	
	\textit{keywords}: astrobots, coordination, convergence prediction, multi-object spectrographs, massive spectroscopic surveys
	\doublespacing
			\section{Introduction}
		\label{sect:intro}
		The nature of dark energy\cite{copeland2006dynamics} is yet unknown to physics. The current mathematical candidates to model dark energy and its dynamics have been remained speculative in view of compatibility with each other and observations. The modern trend of dark energy studies takes an observational standpoint in understanding this concept\cite{zhao2017dynamical}. In particular, one seeks the generation of the map of the observable universe through which the distribution of dark energy may be found. To do so, the light rays emitted from cosmological objects have to be captured by optical fibers mounted on particular ground telescopes. Then, those signals are sent to a spectrograph which synthesizes the local survey of an observation. The overall combination of all local surveys gives rise to the final survey in the course of thousands of observations.
		
\begin{figure}
			\centering\includegraphics[scale=0.4]{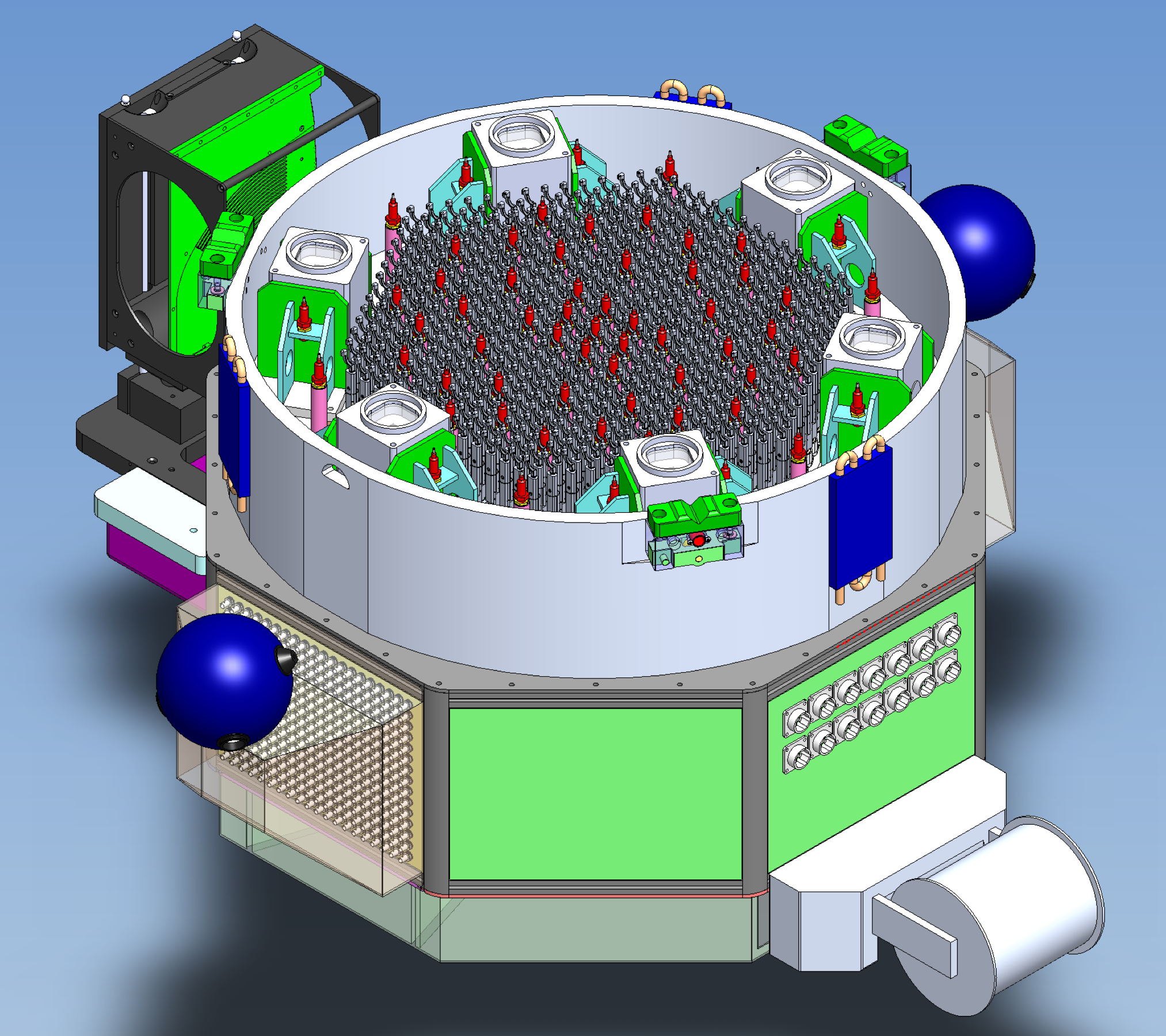}
			\caption{A schematic of an astrobot swarm (reprinted from the SDSS-V wiki with permission)\label{fig:fps}}
		\end{figure}

		Each observation includes a set of unique targets whose projected locations on the focal plane of a typical telescope are different. So, fibers have to be coordinated from one observation to another based on pre-defined target-to-fiber assignments\cite{macktoobian2020optimal}. Since a telescope may encompass hundreds to thousands of fibers, manual coordination of fibers are infeasible because they are both tedious and inefficiently slow. Namely, the available time between two consecutive observations is not long enough to manually reconfigure many fibers. Thus, astrobots\cite{macktoobian2021astrobotics,horler2018robotic} were designed to automatize the coordination of fibers. Each astrobot is a two-degree-of-freedom manipulator, as shown in Fig. \ref{fig:pos}, through which a fiber is passed. The tip of the fiber is located at the end-effector of its hosting astrobot. So, the fiber tip can reach any point belonging to the working space of its astrobot. The area available in the focal plane of a telescope is limited. Thus, astrobots are placed in dense hexagonal formations according to which each astrobot is subject to potential collisions with its neighbors\footnote{\z{A neighborhood associated with an astrobot is defined as the union of the astrobot and its immediate surrounding peers.}}. In particular, Fig. \ref{fig:fps} illustrates a schematic of an astrobot swarm at the back of which a spectrograph is located. The more astrobots are coordinated, the more data are obtained associated with their observation, thereby the higher resolution their resulting map has. Thus, reaching a minimum of convergence rate in a swarm is an important requirement. This complicated swarm control problem is already solved using artificial potential fields \cite{macktoobian2019complete,macktoobian2020experimental,tao2018priority} and supervisory control \cite{macktoobian2019supervisory}. These methods can check the completeness of a typical swarm configuration. \z{The artificial-potential-based methods assign a distributed controller to each astrobot. Each controller generates trajectories by which its corresponding astrobot tends to its target point without colliding with any other peer}. If the completeness\footnote{\z{Completeness refers to the total successful coordination of astrobots so that all of them reach their targets associated with an observation.}} is not fulfilled, one has to search for another configuration and iteratively check its completeness until a completely-reconfigurable configuration is found. The methods above can only be assessed using extensive time-consuming simulations. Thus, the idea of predicting the convergence rate of a swarm is taken into account using the data of its former coordination. The current modern astrobot swarms constitute $\sim$500 units \cite{kollmeier2017sdss} in which the prediction capabilities may save many trial-and-error efforts in their coordination phases. Interestingly, the next-generation swarms will include $\sim$20,000 astrobots\cite{schlegel2019astro2020} for which convergence predictions would be necessary.
		
		\y{The convergence prediction using machine learning seeks an estimation of the post-coordination status of a swarm in view of those targets which can be reached. In this framework, we only take the initial configuration of astrobots without engaging with the intricacies of their interactions in the course of their coordination. Finally, if the estimated prediction is below a desired threshold, one simply switches to another plan of target-to-fiber assignments. The first solution to this problem\cite{macktoobian2020data} proposed a weighted $k$-NN-based algorithm\cite{dubey2013class} to predict the intended convergences. Despite the promising accuracy obtained by this algorithm, it has some drawbacks. In particular, this algorithm is a lazy-evaluating method which does not generate any model of its learning process. Thus, all geometrical computations corresponding to every training data sample have to be done associated with every observation. This computational issue may significantly slow predictions. }
		
		\y{Categorical data cannot be properly embedded into the $k$-NN-based scheme because this technique uses a distance metric which is not applicable to categorical data in a straightforward manner. So, the $k$-NN-based strategy only considers a constrained version of the prediction problem in which only spatial features of the locations of targets are taken into account. However, a more realistic and more accurate model may be trained if one takes the critical categorical features of astrobots, such as parity, as well. Namely, parity denotes the rotation direction of the outer arm of an astrobot. The constrained case solved by the $k$-NN-based approach assumed that the parities of all astrobots are the same. This assumption is so restrictive in terms of decreasing the maneuverability of astrobots. In the sequel, swarm controllers may end up with very low convergence rates because of such extremely restrictive constraint. On the other hand, relaxation of the fixed-parity assumption makes a prediction substantially more difficult because even if one only toggles the parity of one astrobot of a swarm, the convergence of many astrobots may be affected. In other words, the consideration of parity has to be efficiently managed to reach high prediction accuracies. Accordingly, we seek a solution, based on the idea of support vector machines (SVM)\cite{suykens1999least}, to solve a generalized version of the convergence prediction problem in which categorical data, particularly parity, can also be incorporated into prediction processes.}
		\subsection{Contributions}
		The contributions of this paper are three-fold as follows. 
		\begin{itemize}
			\item We obtain a prediction model which can be simply evaluated for arbitrary coordination cases. Put differently, once our model is computed according to a particular swarm, any test scenario associated with that swarm can be instantly evaluated. It is a notable advancement compared to the lazy evaluations of constrained scenarios handled by the $k$-NN-based method. Our SVM-based algorithm outperforms the $k$-NN-based strategy in view of prediction performance. 
			\item The $k$-NN-based scheme requires a neighborhood analysis step to localize distance measurements in the course of its evaluations. However, our SVM-based predictor models each astrobot such that computations are inherently localized. So, one needs no extra pre-processing to localize data before any learning phase. 
			\item We incorporate parity in our algorithm using a normalization phase. In particular, we transform the categorical parity pair to a numerical one whose variation resembles those of the spatial features of astrobots. Thanks to this formulation, no feature dominates the other ones in the learning process.
		\end{itemize}
		\subsection{Organization}
		\begin{figure}
			\centering\includegraphics[scale=0.7]{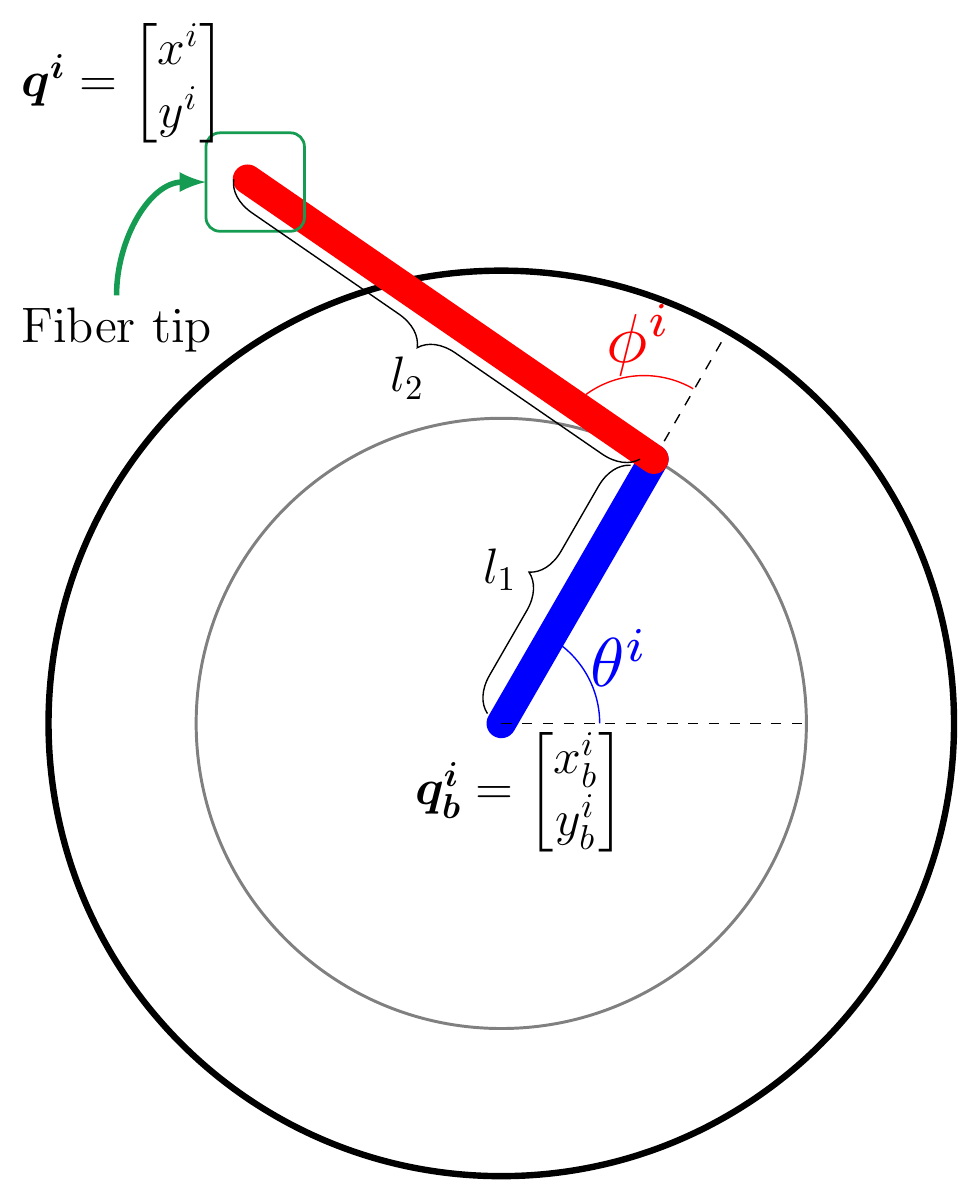}
			\caption{A top view of a typical astrobot (reprinted with permission\cite{macktoobian2019navigation})\label{fig:pos}}
		\end{figure}
		\begin{figure}
			\centering\includegraphics[scale=1.5]{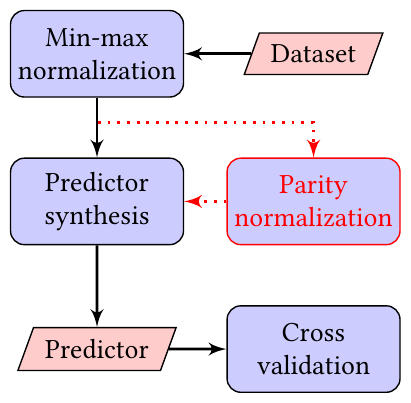}
			\caption{The flowchart of the SVM-based prediction algorithm (the solid black pathway refers the synthesis process for constrained scenarios, while the one including the parity normalization block has to be taken into account for the case of generalized scenarios.) \label{fig:alg}}
		\end{figure}
		\z{Section \ref{subsec:char} reviews the characterizations of astrobots, say, their geometry and kinematics. We then illustrate assumptions regarding our focal plane setup, astrobots specifications, and the data definitions which are used in our formalism and in the course of our simulations, in Section \ref{subsec:general}.} Section \ref{sec:cons} describes the synthesis of our SVM-based predictor in the constrained case, say, when all astrobots have the same parity. The black processing path of Fig. \ref{fig:alg} represents the underlying steps of the constrained case. To be specific, we first define the computational model of an astrobot, including its selected features, which is later used in the learning phase of our algorithm. We then describe the necessity of scaling the features of the astrobot model. A detailed treatment of the predictor synthesis process and the applied cross-validation procedure are also discussed. Section \ref{sec:gen} illustrates how the constrained algorithm can be extended to a generalized one to cover heterogeneous parities in a focal plane, as well. Following the dotted red processing path of the algorithm in Fig. \ref{fig:alg}, once parity is normalized, a predictor can be synthesized according to the formalism used to solve the constrained case. We apply our method to a 487-astrobot swarm to illustrate the efficiency of the synthesized predictor, in Section \ref{sec:res}. Finally, Section \ref{sec:conc} highlights our conclusion.
		\begin{figure}
			\centering\includegraphics[scale=1.3]{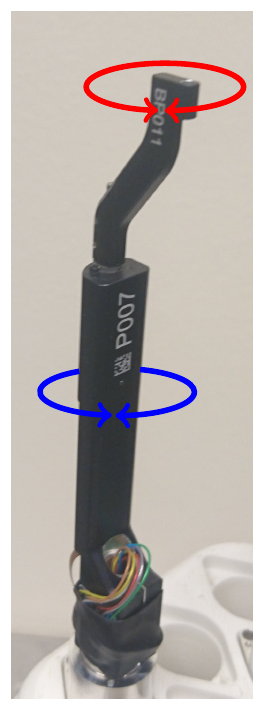}
			\caption{The rotational movements of the arms of a typical astrobot. The red rotation reflects the variable parity of the astrobots since it can rotate in both directions (reprinted with permission\cite{macktoobian2021astrobotics})}
			\label{fig:parity}
		\end{figure}
		\section{Preliminaries}
		\subsection{Astrobot characterization}
		\label{subsec:char}
		\z{We describe the functionality of an astrobot as well as and its kinematic formulation. Each astrobot is a planar rotational-rotational 2-DoF manipulator whose ferrule can move in its working space to reach the point at which the target assigned to its fiber is projected. Accordingly, the forward kinematics of astrobot $i$ may be written as below (see, Fig. \ref{fig:pos})
		\begin{equation}\label{eq:kin}
			\bm{q^{i}} = \bm{q^{i}_{b}} + 
			\bracketMatrixstack{\cos(\theta^{i}) & \cos(\theta^{i} + \phi^{i})\\
				\sin(\theta^{i}) & \sin(\theta^{i} + \phi^{i})} \bm{l},
		\end{equation}
		where the ferrule coordinate of the astrobot is $\bm{q^{i}} = \bracketMatrixstack{x^{i} & y^{i}}^\intercal$ with respect to a universal frame attached to its
		focal plane. $\bm{q^{i}_{b}}= \bracketMatrixstack{x^{i}_{b} & y^{i}_{b}}^\intercal$ denotes the base coordinate of the astrobot. The lengths of its rotational arms are represented by $\bm{l} = \bracketMatrixstack{l_{1} &l_{2}}^\intercal$. The angular deviations of the first and the second arms of the astrobot are represented by $\theta^{i}$ and $\phi^{i}$, respectively.}
	
		\z{Each optical fiber passes through an astrobot so that the tip of the fiber, known as ferrule, is placed at the end-effector area of its astrobot. In the course of an observation, each fiber collects the light associated with its target. Then, a spectrograph processes those signals to generate a survey corresponding to an observation.} 
		\subsection{General data specification}
		\label{subsec:general}
\begin{figure}
			\centering\includegraphics[scale=3]{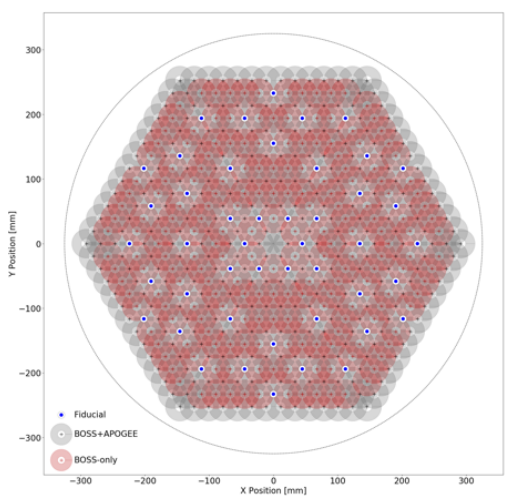}
			\caption{The 2.5m focal plane of SDSS-V the arrangement of whose astrobots are taken into account to generate the coordinates of astrobots in this paper (reprinted with permission\cite{pogge2020robotic}.)}
			\label{fig:foc25}
		\end{figure}

			\z{To generate the data associated with astrobots, we take the 2.5m focal plane of SDSS-V into account \cite{gunn20062}, as depicted in Fig. \ref{fig:foc25}. Thus, the base coordinate associated with each astrobot is located at one of the astrobot slots of this focal plane. The focal plane can host a maximum of 500 astrobots. However, we consider 487 astrobots in our analyses for the purpose of analyzing the difference between convergence prediction in complete neighborhoods and incomplete ones. In particular, a total neighborhood is set of 6 astrobots around a central one in a hexagonal formation, while a partial neighborhood lacks at last one of those circumferential astrobots. In this setting, the length of the first (resp., second) arm of each astrobot is 7.4mm (resp., 15 mm). Thus, the overall pitch is 22.4 mm which exactly equals the distance between the base coordinates of each pair of adjacent astrobots. We assume that the rotational step size of each astrobot is 0.1$^\circ$, and its temporal step size is 0.25 s.}
			
			\z{Each astrobot has to be assigned to a target. We take a uniformly distributed set of targets in polar coordinate system $(r,\theta)$, as below, whose center is located at the center of our focal plane.
				\begin{equation}
					\begin{split}
						&\theta \sim \mathcal{U}[-\pi;\pi]\\
						&r^2 \sim \mathcal{U}(0;r^{2}_{\mathrm{max}})
					\end{split}
				\end{equation}
				Here,  $r_{\mathrm{max}}$ denotes the radius of the focal plane. Targets are uniformly selected from an astrobot's x-y workspace. Thus, the reachability is automatically checked at the assignment stage. We conduct the assignment, subject to the fulfillment of reachability requirement, in a random manner to remove any bias regarding the usage any particular assignment method.} 
				
				\z{An important parameter of a typical astrobot is parity. Parity determines the rotation direction of the second arm of an astrobot. By convention, parity 0 (resp., 1) refers to the clockwise (resp., counterclockwise) motions of the cited arm. In this paper, we study the convergence prediction of astrobots in two scenarios with respect to parity. First, we take a constrained case into account in which the parities of all astrobots are homogeneously either 0 or 1. In this case, we expect that our prediction procedure yields relatively accurate prediction because astrobots motions are constrained. However, this potential ease of prediction comes at the cost of the shrinkage of the potential trajectories astrobots may find to reach their targets. In such constrained cases, we randomly assign a fixed parity to each astrobot. On the other hand, we also consider another scenario in which astrobots can move in both directions from one observation to another one, that is, their parities are heterogeneous. The generalized setting described above may provide more degrees of freedom to controllers of astrobots to plan trajectories for them. However, accuracy of convergence predictions may be slightly worse than those of the constrained case.} 
		
\begin{figure}
			\centering\includegraphics[scale=0.7]{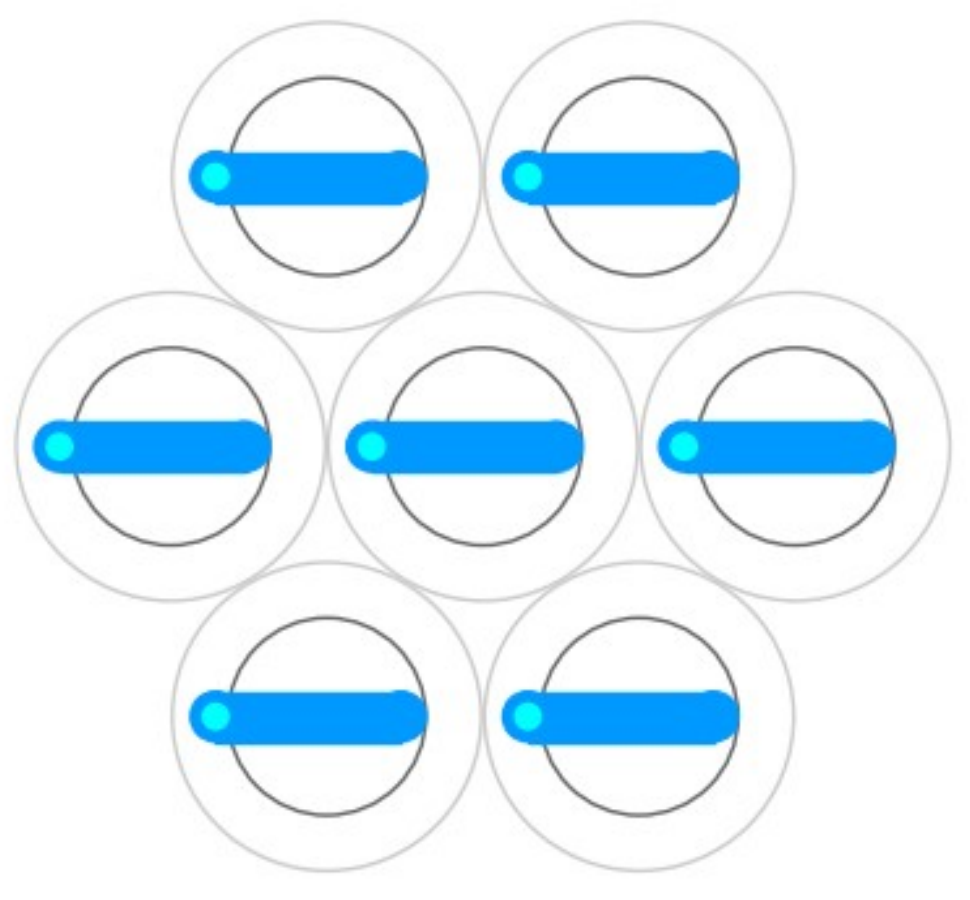}
			\caption{The folded formation of astrobots which is assumed to be the initial formation in the course of our prediction synthesis processes}
			\label{fig:full}
		\end{figure}

		\z{The path generator we use\cite{tao2018priority} to generate these data seeks a path for astrobots beginning from an initial "folded formation" (as shown in Fig. \ref{fig:full}), to a desired "target formation" (in which astrobots may have any formation). Our algorithm behaves similarly to an N-body simulation where astrobots are attracted to their targets and repelled from their neighbors. When applied to SDSS-style robots, many astrobots do not achieve their targets, due to their extreme probability of interference, etc. We wish to predict the success of this path generator without having to run it, which will provide a method for assessments for target selection. This is useful because the complete path generator \cite{macktoobian2019complete,macktoobian2020experimental} may require some time longer than the available time between two consecutive observations to run, which makes it computationally expensive for path generations. So, we are interested in using those which do not generally fulfill completeness, should they reach a minimum satisfactory convergence rate.}
			
		\z{We collect data by repeating many independent simulations (10100 and 15100 iterations for constrained and generalized cases, respectively). After each simulation, we discover which astrobots achieved their targets and which astrobots did not. We use this information to build an SVM model to predict which astrobots will ultimately be successful in achieving targets given target assignments without the need of running this expensive computation. This will be most relevant in target-assignment phases of a survey where the repeated running of our path generator may not be computationally feasible.}

		\z{In the next sections, we use the data associated with many former coordination of a set of astrobots to train a predictor. This predictor functionally resembles a function: given the set of astrobots and a new observation, one feeds the pairings of the astrobots and their targets, as well as their (fixed or variable) parities into this function. The function then returns a 1 (resp., 0) flag for each astrobot indicating whether the astrobot's convergence to its target is predicted to be successful (resp., unsuccessful). If the overall prediction is satisfactory, one may run the path generator to obtain the trajectories of the astrobots for that observation. Otherwise, new pairings may be taken into account until the predictor finds the one which satisfies a desired minimum convergence rate. In the course of applying cross validation to our model, we shuffle each dataset then randomly partition its data based on what our cross-validation process requires.}
		\section{Constrained convergence prediction \protect\footnote{Throughout this paper, scalars and boolean parameters are represented by regular symbols. Bold symbols are reserved to denote (sets of) matrices.}}
		\label{sec:cons}
		\subsection{Data Definition}
		In the constrained scenario, we assume that the parities of all astrobots are fixed and similar to each other. Since each coordination starts from the folded formation of astrobots, similarly to the $k$-NN-based algorithm, the data construction associated with the coordinate of each astrobot corresponds to the coordinate of the projected location of the particular target assigned to it. Thus, given a fixed parity for all astrobots of a swarm, the constrained data model of astrobot $\pi$ that has $n$ neighbors is defined as a collection of spatial features as follows\footnote{Operator $(\cdot)^\intercal$ yields the transpose of its matrix argument.}
		\begin{equation}
		\label{eq:cons}
		\bm{\pi^{\mathrm{C}}} \coloneqq \begin{bmatrix}
		x_{t} & y_{t} & x^{1}_{t} & y_{t}^{1} & \cdots & x^{n}_{t} & y_{t}^{n}
		\end{bmatrix}^\intercal.
		\end{equation}
		Here $\begin{bmatrix}x_{t} & y_{t}\end{bmatrix}^\intercal$ denotes the target coordinate of $\pi$, and each vector $\begin{bmatrix}x^{i}_{t} & y^{i}_{t}\end{bmatrix}^\intercal$ is associated with the target coordinate of its neighbor $i$. One notes the absence of any parity information in the data vector because it is fixed for all astrobots. Note that the model above is basically including only one neighborhood. Thus, one need not localize the data before being fed into any learning process. In our strategy, given a data vector $\bm{\pi^\mathrm{C}}$, our predictor exclusively returns its evaluation regarding the convergence of the central astrobot $\pi$. So, to predict the convergence of $n$ astrobots of a swarm, one has to take $n$ neighborhoods into account the central astrobot of each of which is one of those $n$ astrobots.
		\subsection{Feature Scaling}
		Before using data vectors to train any predictor, we have to scale the features of each data vector. Namely, target’s coordinates of different neighbors do not vary in similar ranges for all data vectors. So, there is an intrinsic spatial offset in the features with respect to a particular astrobot. Such an offset has to be removed not to synthesize biased predictors. Lack of feature scaling in the case of SVM-based predictors makes the setting of their hyperparameters very complicated. Comparatively, the $k$-NN-based algorithm does not require any feature scaling since the Euclidean distance metric applied to that method is relative and localized. In contrast, we will later see that the quoted metric is used in the Gaussian kernel of our SVM-based predictor. So, we have to take feature scaling into account. In particular, we use min-max normalization \cite{han2011data} to linearly transform the range of all data features to the interval [-1,1]. This range not only removes the mean value of each feature but also yields satisfying results in view of the performance of synthesized predictors. Mathematically, the following formalism maps feature $x$ to its normalized counterpart $x'$ which varies in the cited range.
		\begin{equation}
		x' \coloneqq \frac{2(x-\min{(x)})}{\max{(x)} - \min{(x)}}-1 
		\end{equation}
		Here, operators $\min{(x)}$ and $\max{(x)}$ return the minimum and the maximum values of the feature $x$, respectively, associated with a data vector $\bm{\pi^\mathrm{C}}$ of a particular dataset.
		
		One may note that the linear nature of the transformation above indeed preserves relative distances between the targets of a particular neighbohrood. We generally assume a uniform distribution of targets all over a focal plane. Thus, min-max normalization is a better option compared to Z-score normalization\cite{jain2005score} which is often applied to the data following Gaussian distributions.
		\subsection{Predictor Synthesis}
		It is unlikely that a linear boundary can generally solve the convergence prediction problem of an astrobot swarm. Thus, we apply the kernel trick \cite{hofmann2006support} to our linearly inseparable data. In particular, we map data vector $\bm{\pi^{\mathrm{C}}}_{i}$ to $\bm{\pi^{\mathrm{C^{'}}}}_{i}$ by the following kernel $\kappa(\cdot,\cdot)$, known as radial basis function \cite{murphy2012machine},
		\begin{equation}
		\label{eq:minmax}
		\kappa(\bm{\pi^{\mathrm{C}}}_{i},\bm{\pi^{\mathrm{C^{'}}}}_{i}) := \phi(\bm{\pi^{\mathrm{C}}}_{i})^\intercal\phi(\bm{\pi^{\mathrm{C^{'}}}}_{i}) = \mathrm{exp} \left(-\frac{\norm{\bm{\pi^{\mathrm{C}}}_{i}-\bm{\pi^{\mathrm{C^{'}}}}_{i}}^2}{2\sigma^2}\right).
		\end{equation}
		Here, the kernel size $\sigma$ determines the width of the Gaussian kernel.
		
		The problem of predictor synthesis is equivalent to the solution of the optimization problem below. We seek an optimal hyperplane which classifies the convergence of a particular set of astrobots into the class of 1s (resp., 0s) if they are predicted to reach their targets (resp., otherwise).
		\begin{equation}
		\label{eq:prob}
		\begin{aligned}
		\min_{\bm{w},b,\xi} \quad & \frac{1}{2}\bm{w}^\intercal\bm{w}+\mathcal{C}_{0}\sum_{i=1}^{N_{0}}{\xi^{0}_{i}}+\mathcal{C}_{1}\sum_{i=1}^{N_{1}}{\xi^{1}_{i}}\\
		\textrm{s.t.} \quad & g^{j}_{i}(\bm{w}\phi(\bm{\pi^{\mathrm{C}}}_{i})+b) \geq 1-\xi^{j}_{i}, &\quad j\in \{0,1\}\\
		&\xi^{j}_{i} \geq 0, &\quad j\in \{0,1\}   \\
		\end{aligned}
		\end{equation}
		The boundary between the two classes is denoted by normal vector $\bm{w}$. $N_0$ and $N_1$ are the numbers of the samples in the classes of 0s and 1s, respectively. Weights $\mathcal{C}_{0}$ and $\mathcal{C}_{1}$ represent the miss-classification penalties associated with the classes of 0s and 1s, respectively. Given miss-classification measure $C$, we have
		\begin{equation}
		C_{j} := C\omega_{j}, \quad j\in \{0,1\},
		\end{equation}
		where $\omega_{j}$ is the class weight of class $j$. The notion of class weight is also used to resolve the imbalanced data problem. To balance the bias with respect to the abundance of the majority class, i.e., 1s, compared to the minority class, i.e., 0s, we apply the class weights to the class of 1s. For this purpose, one has to regulate the hyperparameter $\omega_{1}$. In general, one may safely apply either a smaller weight to the majority class or a larger one to the minority class. The quantities $\xi_{i}^{0}$ and $\xi_{i}^{1}$ are the slack variables corresponding to the incorrect classifications of samples $i$ regarding the classes of 0s and 1s, respectively. Due to the complexity of our prediction problem, we use these variables to relax classification constraints and allow miss-classifications of some data samples. These values are larger than 0 only if their corresponding samples are miss-classified. Moreover, the ground truth of sample $i$ is encoded by complement-pair $(g_{i}^{0}, g_{i}^{1})$ with respect to the classes of 0s and 1s, respectively. Astrobot $i$ is represented by $\bm{\pi^{\mathrm{C}}}_{i}$, and  kernel function $\phi(\bm{\pi}_{i})$ maps every feature of  $\bm{\pi}_{i}$ into a higher dimensional space. $b$ indicates the hyperplane intercept. One observes that the term $\frac{1}{2}\bm{w}^\intercal\bm{w}$ is the converse of the margin between the two classes. The minimization of this term indeed gives rise to the maximization of the desired margin corresponding to the predictor boundary. The setting of the cited hyperparameters are described in Section \ref{sec:res}.
		
		The optimization problem (\ref{eq:prob}) is solved using the sequential minimal optimization algorithm \cite{platt1998sequential}. The obtained boundary hyperplane is computed according to a subset of the data samples, i.e., support vectors, which are the closest data points to the hyperplane. This linear hyperplane, in the expanded space, is projected back to the original nonlinear space. Once the hyperplane is found, we assign a new test vector to one of the two classes of the problem. This assignment depends on the relative position of the data vector with respect to the hyperplane model.
		\subsection{Validation}
		We employ $k$-fold cross validation method to check the performance of our algorithm. Namely, we synthesize a desired SVM-based convergence predictor using a train partition $D_T$ of a particular dataset $D$. Then, we apply the algorithm to the second partition, say, a test partition $D_S$, to assess the algorithm's performance. In this regard, we perform\footnote{Operators $\lvert \cdot \rvert$ and $\floor*{\cdot}$ return the cardinality and the floor of their arguments, respectively.} $\floor*{\lvert D\rvert / \lvert D_{S}\rvert}$ validation iterations. Thanks to this method, all elements of a dataset are used both as a part of the train and test partitions in the end of a cross-validation process. Moreover, one may not use one data sample more than once, as it may happen in the Monte-Carlo cross validation\cite{xu2001monte}. In each iteration, we take a different partition of $D$ as $D_S$. We compute the average of the performance results obtained at the end of every iteration over the total number of $k$ iterations, as the final results of the cross-validation process. The value of $k$ depends on the ratio $\floor*{\lvert D\rvert / \lvert D_{S}\rvert}$. Decreasing the cardinality of $D_S$ increases the value of $k$ and the required time for the completion of the cross-validation process. However, increasing the size of $D_S$ implies the usage of less samples in the training phase. Thus, less training data may escalate the risk of underfitting in the course of the SVM-based predictor synthesis. Section \ref{sec:res} illustrates our setting corresponding to this hyperparameter.
		\begin{figure}
				\centering\includegraphics[scale=0.3]{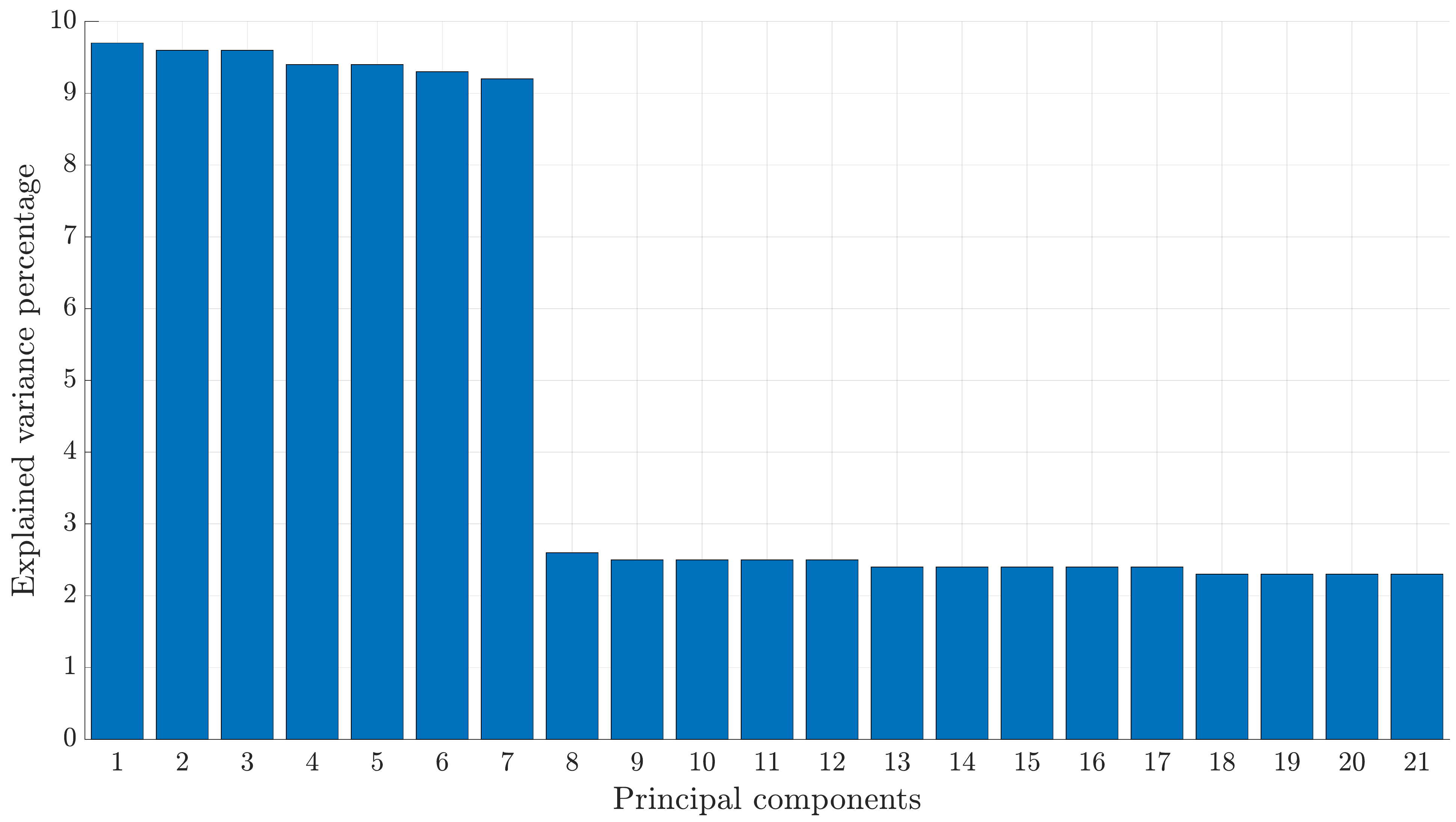}
				\caption{Unbalanced explained variance before parity normalization\label{fig:notscaled}}
		\end{figure}
		\begin{figure}
			\centering\includegraphics[scale=0.3]{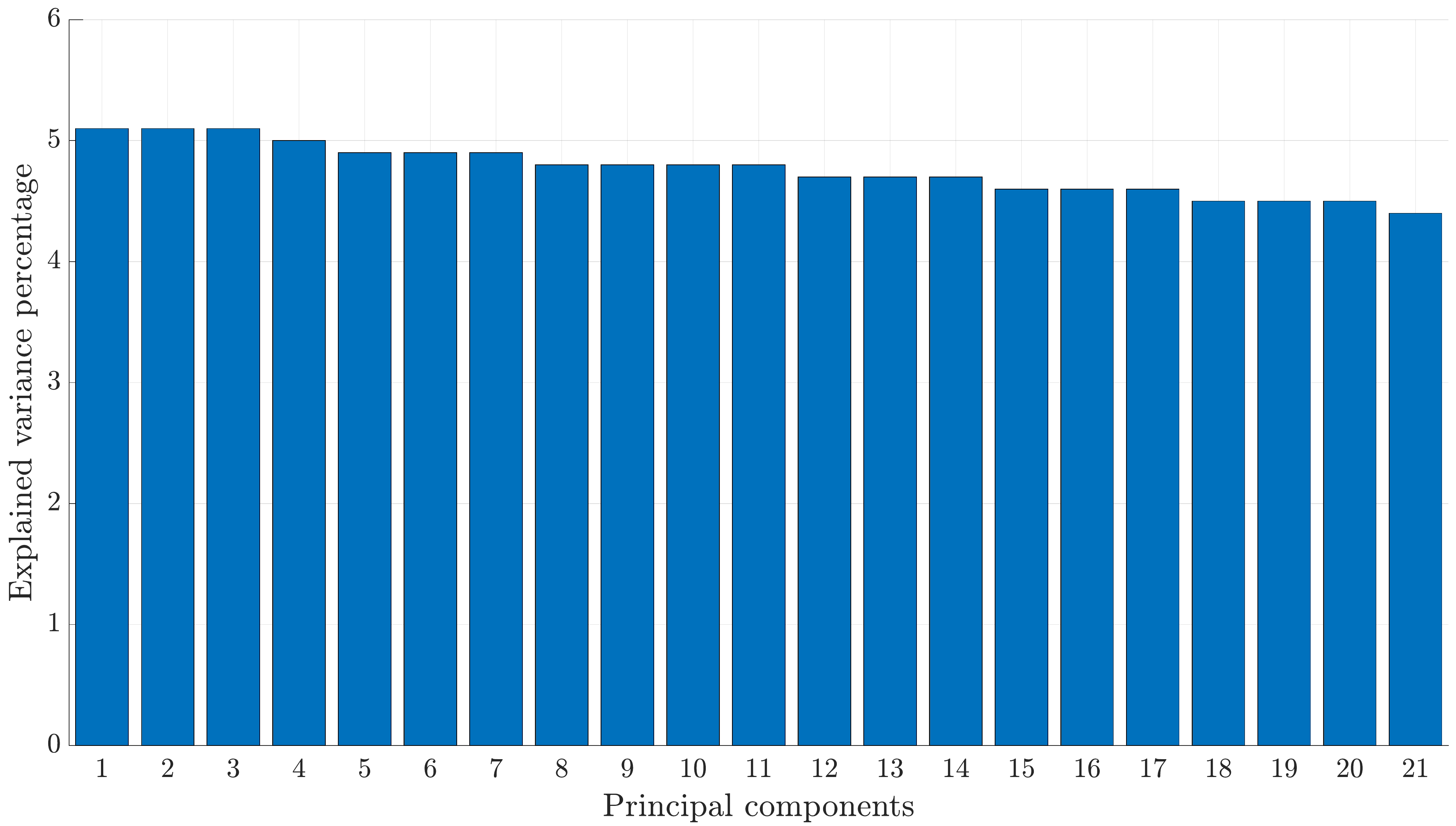}
			\caption{Balanced explained variance after parity normalization\label{fig:scaled}}
		\end{figure}
		\section{Generalized convergence prediction}
		\label{sec:gen}
		In this section, we generalize our convergence prediction by adding a parity to the features of each astrobot. Put differently, according to this generalization, each astrobot's parity may be different from those of other astrobots. This assumption makes the convergence prediction process even more complicated because various parities increase the nonlinear interactions of astrobots. In particular, such interactions give rise to the grow of the potential deadlock and/or collision-prone scenarios. So, predicting the safety and eventual completeness of any swarm initial configuration becomes more challenging. We describe how the generalized version of our SVM-based algorithm efficiently manages to predict the desired safe complete convergences.
		\subsection{Data definition generalization}
		As we have already noted, the notion of parity denotes the rotation direction of rotation of the outer arm of a typical astrobot. Thus, it is inherently classified as a categorical information, contrary to the continuous numerical values corresponding to the remainder of an astrobot's spatial features. In this section, we add parity information to the constrained data vector of an astrobot (\ref{eq:cons}) to obtain the generalized data vector as follows
		\begin{equation}
		\label{eq:gen}
		\bm{\pi^{\mathrm{G}}} \coloneqq \begin{bmatrix}
		x_{t} & y_{t} & P & x^{1}_{t} & y_{t}^{1} & P^{1} & \cdots & x^{n}_{t} & y_{t}^{n} & P^{n}
		\end{bmatrix}^\intercal,
		\end{equation}
		in which $P$ refers to the parity flag of the modeled astrobot $\pi$, and $\{P^{i} \mid 1 \ge i \ge n \}$ denotes the parity set corresponding to $n$ neighbors of $\pi$.
		\begin{figure}
			\centering
			\includegraphics[scale=0.4]{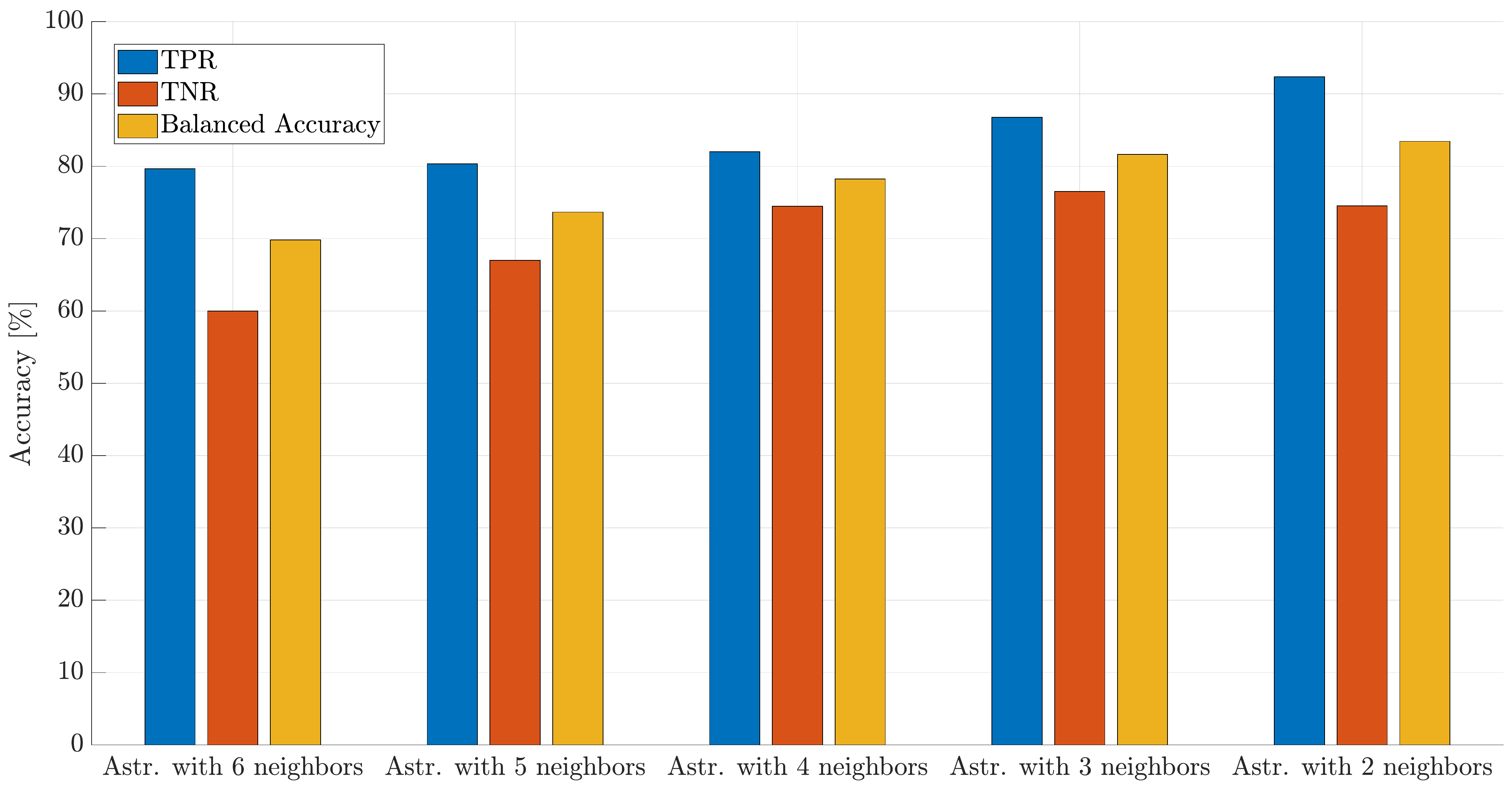}
			\caption{Prediction accuracy per neighborhood type in the constrained scenario}
			\label{fig:anal}
		\end{figure}
		\subsection{Parity normalization}
		In addition to the min-max normalization applied to the spatial features (\ref{eq:minmax}), this section describes the importance of parity normalization, as well. In particular, a categorical parity value is either 1 (resp., -1) to represent clockwise (resp., counterclockwise) rotations of the outer arm of an astrobot. In view of the optimization problem (\ref{eq:prob}), parity data are processed as integer numbers. Thus, the range of their variation has to be normalized such that they vary in a more-or-less similar range as those of the normalized spatial features.
		\begin{figure}
			\centering
			\includegraphics[scale=0.35]{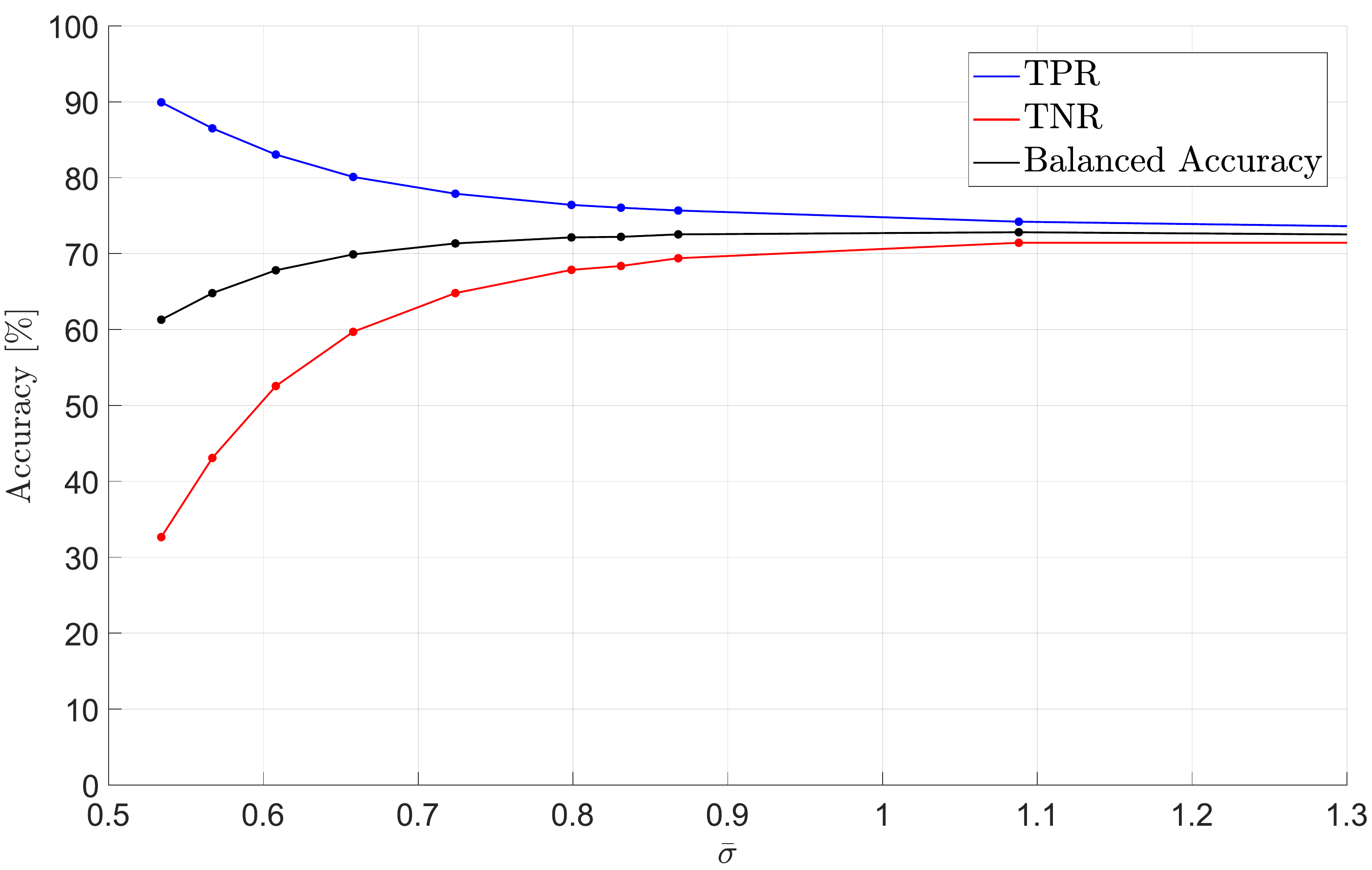}
			\caption{TNR, TPR, and balanced accuracy dynamics with respect to average kernel size in the constrained scenario}
			\label{fig:vs}
		\end{figure}
		
		To efficiently normalize parity, we analyze the standard deviation corresponding to the spatial features. Targets distribution is generated by a standard uniform distribution. However, one cannot simply yield the desired standard deviation using ideal formula associated with standard uniform distribution, which is $\sim$0.577. Since the outer arm of an astrobot is longer than its first arm, the reachability requirement of a target-to-astrobot assignment requires that any generated target may be located outside of the focal plane as long as the target is still reachable by at least one astrobot. The realization of the aforesaid condition empirically requires that one computes the desired spatial standard deviation based on not the general formula but the available data. If we take values -1 and 1 to represent various parities, then the parity standard deviation is $\sim$1. It turns out that this encoding leads to a noticeable imbalance in view of the data explained variance\footnote{Explained variance is the ratio of the variance of a specific feature to the summation of the variances of all features of data.}. In other words, the parity pair (-1,1) makes our SVM-based predictor biased in relying more on the information given by parities compared to those of the spatial features, thereby reducing the prediction quality of the final learning model. 
		\begin{figure}
			\centering
			\includegraphics[scale=0.4]{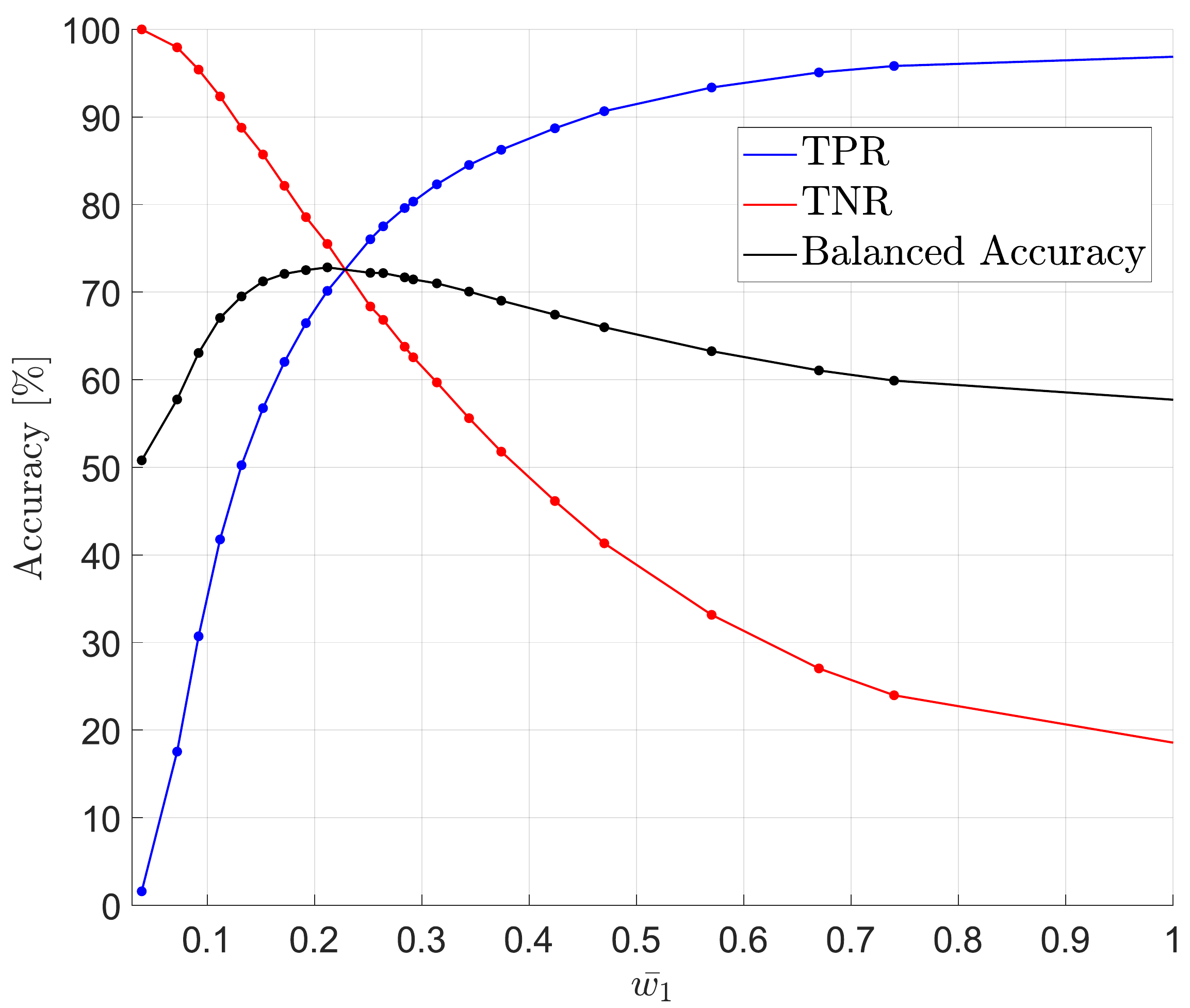}
			\caption{TNR, TPR, and balanced accuracy with respect to average class weight in the constrained scenario}
			\label{fig:omegabl}
		\end{figure}
		
		We obtain the explained variance associated with the features of an astrobot with six neighbors using principal component analysis, as depicted in Fig. \ref{fig:notscaled}. In this figure, the feature imbalance is obvious. So, we need to scale parity values such that their resulting explained variance is fairly similar to that of the spatial features. In particular, the analysis of the target's distributions indicates that the standard deviation of spatial features is $\sim$0.5. On the other hand, feature imbalance issue mandates that the parity standard deviation follows that of the spatial features. One notes that the parity pair (-0.5, 0.5) fulfills the quoted conditions. Taking the planned parity pair into account efficiently scales the explained variance of all features, as Fig.~\ref{fig:scaled} illustrates. We note that choosing parity pairs with a smaller variation range, e.g., (-0.3, 0.3), leads to another issue. In particular, such pairs undesirably increase the impact of the information of spatial features by taking less impact of parities into account on the prediction model synthesis. 
		
		Once parities are normalized, one simply feeds all $\bm{\pi^{\mathrm{G}}}$ vectors to the optimization problem (\ref{eq:prob}). The resulting boundary is the solution to the generalized convergence prediction problem.
		\section{Results\protect\footnote{The simulations are performed on a Dell Inspiron 15 7000 with a processor Intel Core i7-7700HQ, a 2.80 GHz CPU,  16 GB of RAM, run on a Windows 10 Home 64 bit.}}
		\label{sec:res}
		In this section, we demonstrate the performance of our algorithm applied to a 487-astrobot swarm in both constrained and generalized cases. This swarm resembles the one corresponding to the SDSS-V project \cite{kollmeier2017sdss}. We partition our dataset such that every time the test partition is 10\% of the overall dataset, thereby $k=10$. We describe how our algorithm not only solves the constrained case with higher performance compared to the $k$-NN-based algorithm but also efficiently solves the generalized case. We define our atomic performance metrics as below.
		\begin{itemize}
			\item A true positive (TP) is an astrobot which is predicted to converge (the predictor predicts 1), and it actually converges to its target position (its corresponding ground truth element is 1).
			\item A false positive (FP) is an astrobot which is predicted to converge (the predictor predicts 1), but it actually does not converge to its target position (its corresponding ground truth element is 0).
			\item A true negative (TN) is an astrobot which is not predicted to converge (the predictor predicts 0), and it actually does not converge to its target position (its corresponding ground truth element is 0)
			\item A false negative (FN) is an astrobot which is not predicted to converge (the predictor predicts 0), but it actually converges to its target position (its corresponding ground truth element is 1).
		\end{itemize}
		We also define balanced accuracy as the average of TP rate (i.e., TPR) and TN rate (i.e., TNR). 
		
		We synthesize an SVM model for each astrobot of the swarm. After the completion of each iteration, we compute each of the average performance metrics corresponding to each astrobot's atomic performance metric over the number of all iterations. Then, we obtain the performance metrics of the swarm by averaging over those of all astrobots.
		
		We fix $C=1$, so two hyperparameters $\omega_{1}$ and $\sigma$ have to be set for each astrobot. Such setting associated with a particular astrobot critically depends on the cardinality of its neighborhood. Each astrobot empirically possesses two to six astrobots. Thus, the hyperparameter pair above has to be determined for five various scenarios. For this purpose, we tune class weight $\omega_{1}$ and kernel size $\sigma$ for each neighborhood type by performing a grid search.
		\subsection{Constrained Scenario}
		In this scenario, we assume that the parities of all astrobots are the same. Each astrobot dataset comprises 10100 samples. We set the hyperparameters regarding two interesting cases. As Table~\ref{tbl:1} illustrates, case I intends to simultaneously maximize balanced accuracy and keep TPR above 75\%. Case II seeks a minimum TPR of 80\% at the expense of the balanced accuracy decrement. One may note that, given either of the cases, each hyperparameter varies in a relatively narrow margin with respect to a neighborhood type. In other words, the prediction performance is fairly invariant to the variation of the hyperparameters around some particular values. This feature is computationally very important in that one may simply consider a single setting of each hyperparameter for all astrobots of a swarm regardless of differences among their neighborhood types. In this regard, the prediction performance would be sufficiently high, yet no extensive grid search is done to tune hyperparameters based on their specific neighborhood types. The described neighborhood-dependence of prediction accuracies is depicted in Fig. \ref{fig:anal}. 
		\begin{figure}
			\centering
			\includegraphics[scale=0.35]{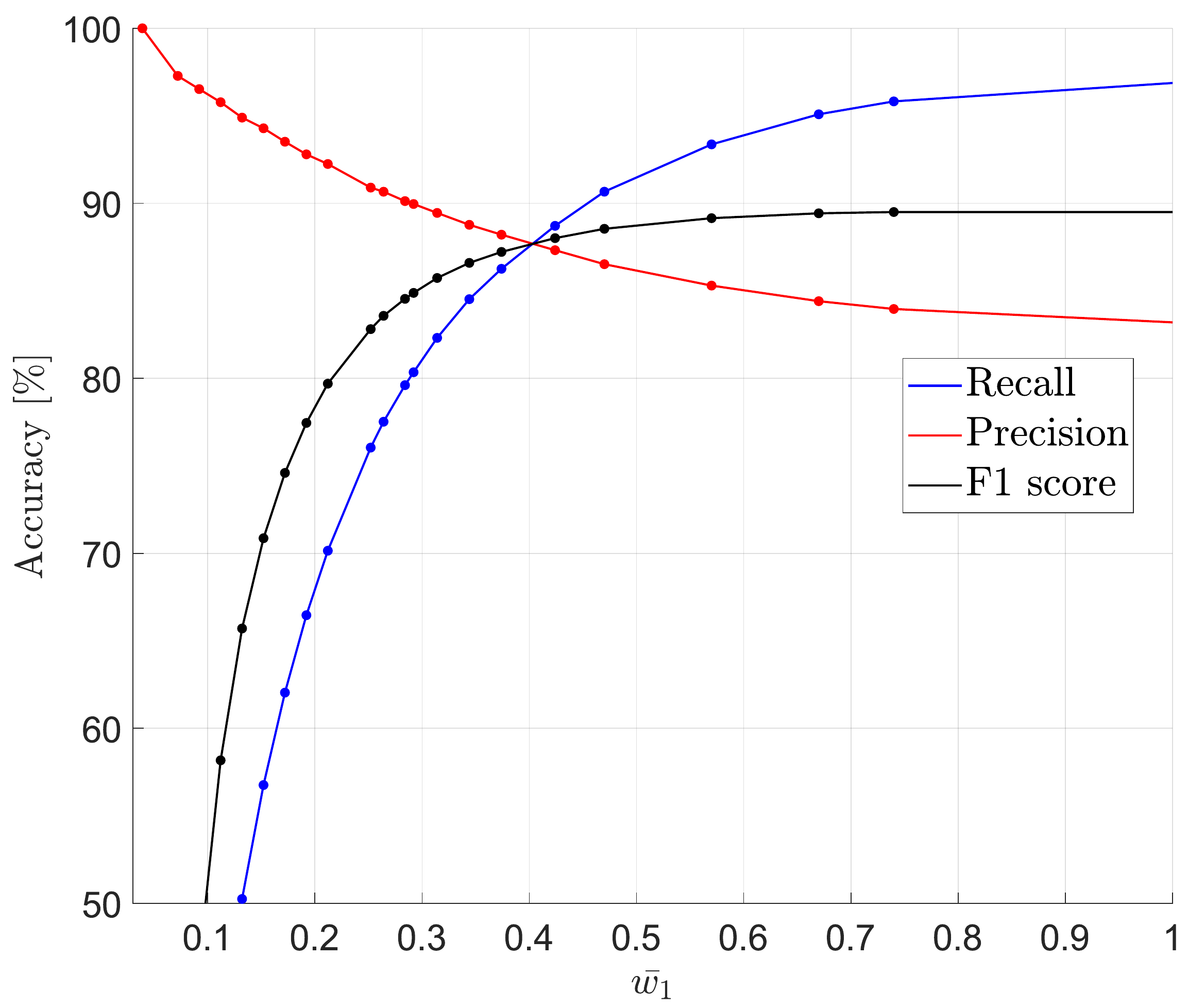}
			\caption{Precision, recall and F1 dynamics with respect to average class weight in the constrained case}
			\label{fig:omegaf1}
		\end{figure}
		
\begin{figure}
			\centering
			\includegraphics[scale=0.35]{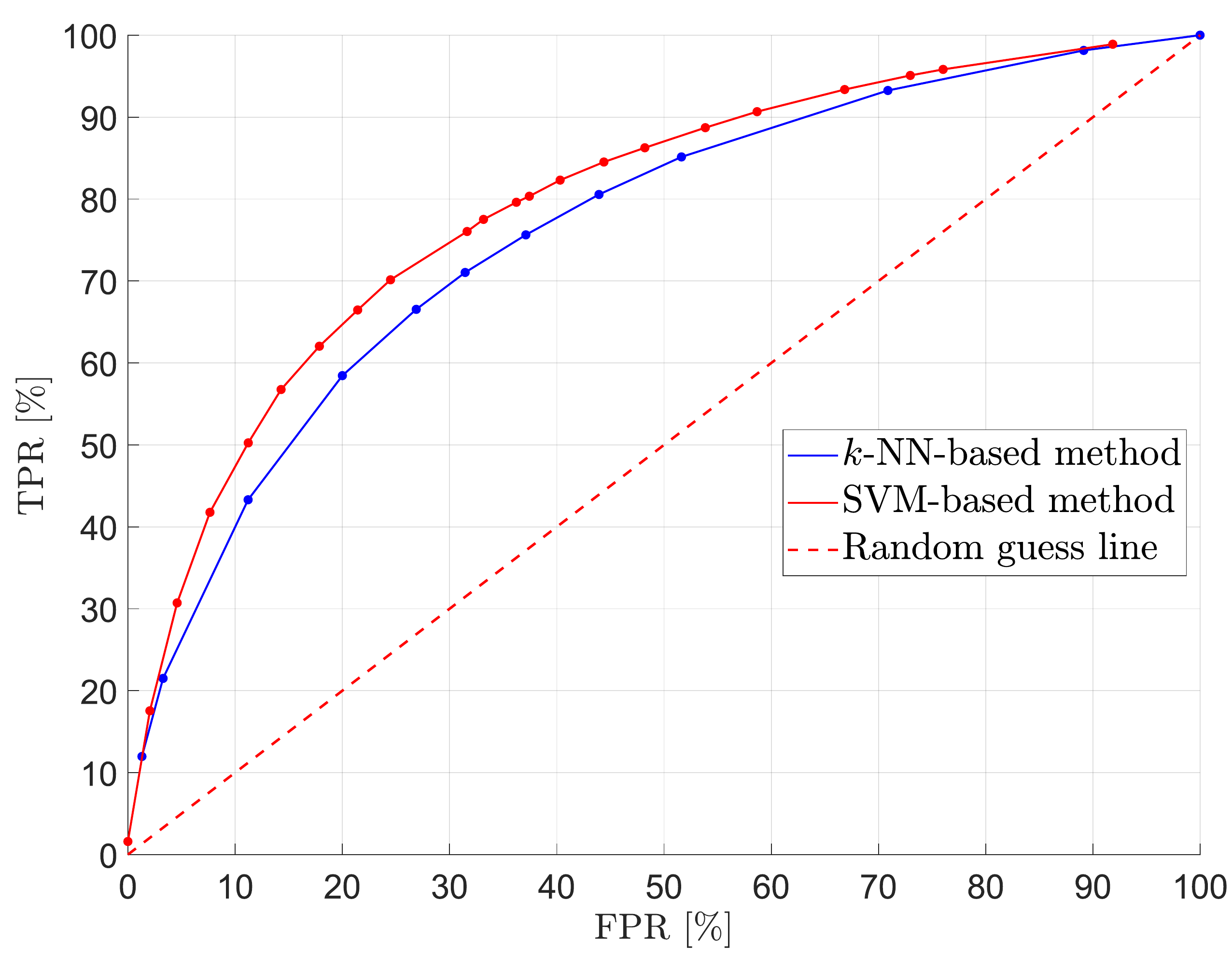}
			\caption{Comparative ROC curves corresponding to the constrained scenario}
			\label{fig:roc}
		\end{figure}

		The impact of the kernel size on the trade-off between TPR and TNR is illustrated in Fig. \ref{fig:vs}. Accordingly, $\sigma \ge 0.8$ provides TPR values over 80\%. The variation trends of balanced accuracy and F1 score with respect to the class weight $\omega_{1}$ are also rendered in Fig. \ref{fig:omegabl} and \ref{fig:omegaf1}, respectively. The optimal trade-off selections are those points at which the graphs intersect. The ROC curve represented in Fig.~\ref{fig:roc} clearly depicts how our SVM-based algorithm is more efficient than the $k$-NN-based one. In particular, the ROC curve of our algorithm is located farther from the random guess line compared to that of the $k$-NN-based one. The overall report of the best prediction performance is reported in Table \ref{tbl:1}.
		\begin{table}
			\centering
			\caption{The best prediction results corresponding to the constrained case (NT refers to neighborhood type which represents the number of the neighbors of an astrobot. BA designates balanced accuracy.)}
			\begin{tabular}{c@{\hskip 0.3in}c@{\hskip 0.3in}c@{\hskip 0.2in}c@{\hskip 0.2in}c@{\hskip 0.2in}c@{\hskip 0.2in}c@{\hskip 0.2in}c@{\hskip 0.2in}c}
				\toprule  
				Case&NT&$\omega_1$&$\sigma$&TPR(\%)&TNR(\%)&BA(\%)&Precision(\%)&F1(\%)\\\cmidrule{1-9}
				\multirow{5}{*}{I}&6&0.277&0.86&\multirow{5}{*}{75.7}&\multirow{5}{*}{69.4}&\multirow{5}{*}{72.5}&\multirow{5}{*}{91.12}&\multirow{5}{*}{82.68}\\
				&5&0.216&0.90&&&&&\\
				&4&0.167&0.86&&&&&\\
				&3&0.167&1.47&&&&&\\
				&2&0.179&0.84&&&&&\\\cmidrule{1-9}
				\multirow{5}{*}{II}&6&0.317&0.82&\multirow{5}{*}{80.3}&\multirow{5}{*}{62.6}&\multirow{5}{*}{71.45}&\multirow{5}{*}{89.96}&\multirow{5}{*}{84.88}\\
				&5&0.256&0.86&&&&&\\
				&4&0.207&0.82&&&&&\\
				&3&0.207&0.98&&&&&\\
				&2&0.219&0.80&&&&&\\
				\bottomrule		
			\end{tabular}
			\label{tbl:1}
		\end{table}
		\subsection{Generalized Scenario}
		We generalize the convergence prediction of the swarm studied in the previous section by randomly determining the parities of its astrobots. Since the generalized case is often more complex than the constrained one, we take 5000 extra samples per astrobot compared to the previous scenario, i.e., 15100 data samples, into account. Table \ref{tbl:2} includes the hyperparameter setting to achieve the best predictions in two cases similar to the constrained case. Namely, the case I and II seeks the maximum balanced accuracy and the maximum TPR, respectively. 
		\begin{figure}
			\centering
			\includegraphics[scale=0.35]{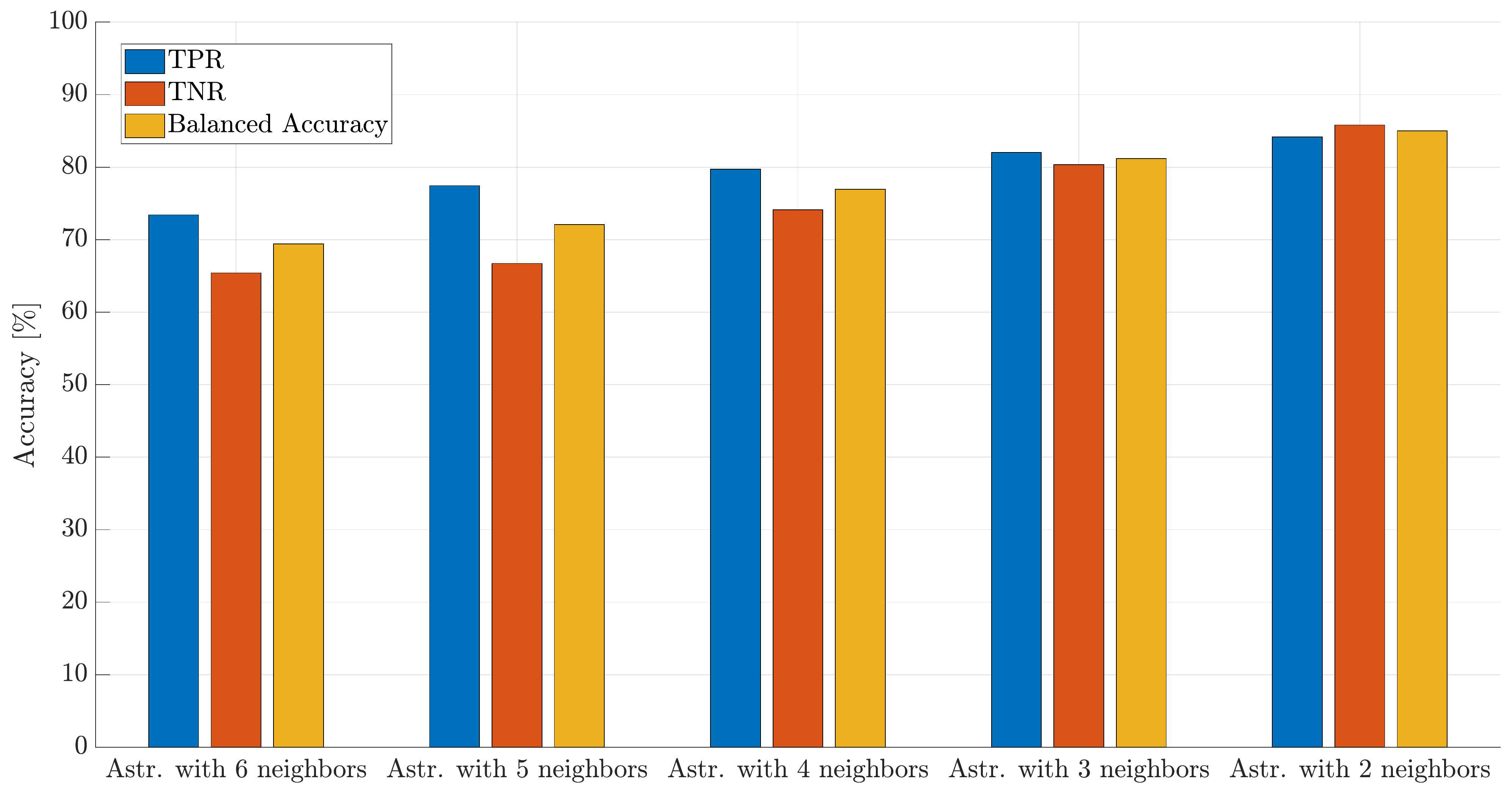}
			\caption{Prediction accuracy per neighborhood type in the generalized scenario}
			\label{fig:analg}
		\end{figure}
		
\begin{figure}
			\centering
			\includegraphics[scale=0.35]{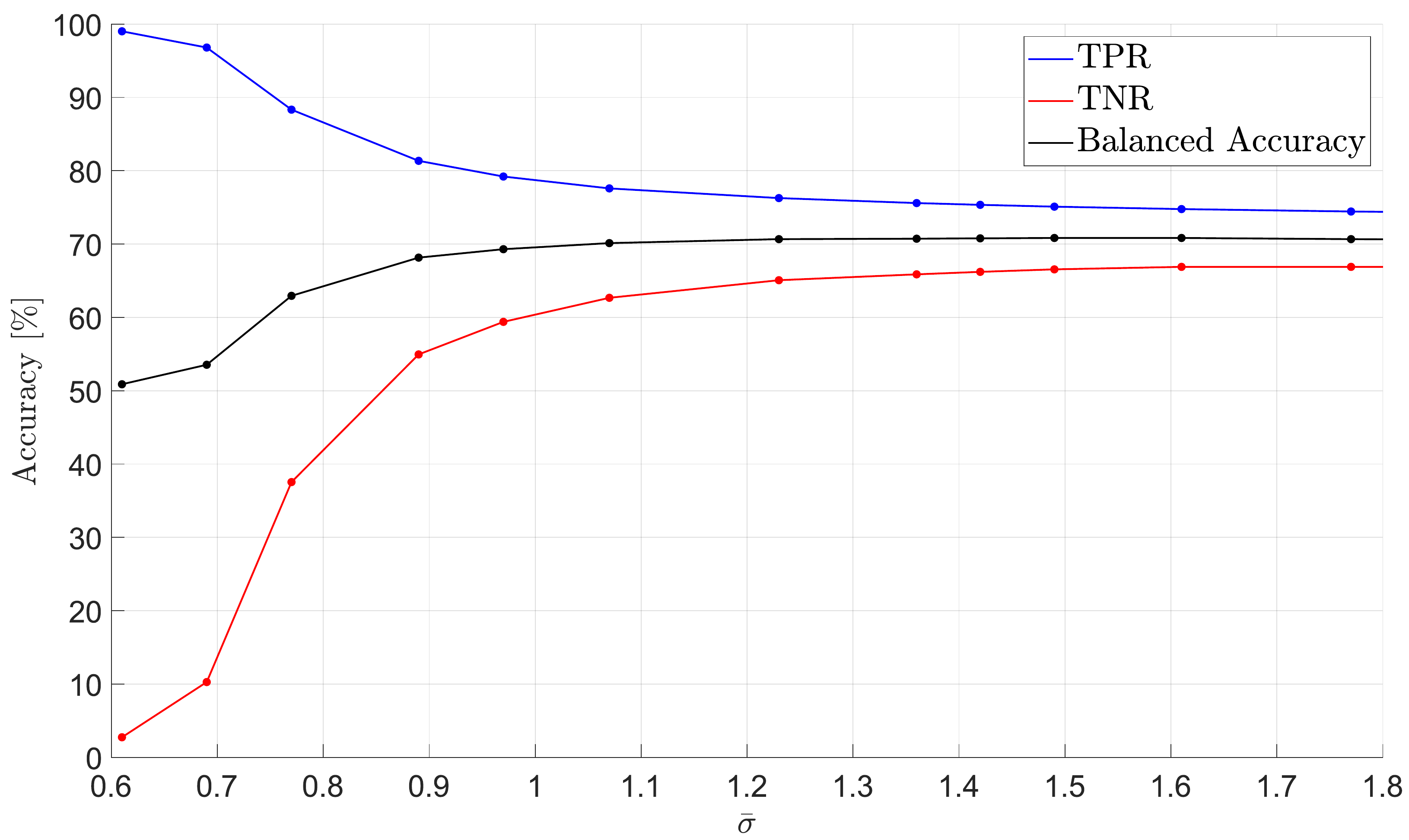}
			\caption{TNR, TPR, and balanced accuracy dynamics with respect to average kernel size in the generalized scenario}
			\label{fig:vsg}
		\end{figure}

\begin{figure}
			\centering
			\includegraphics[scale=0.35]{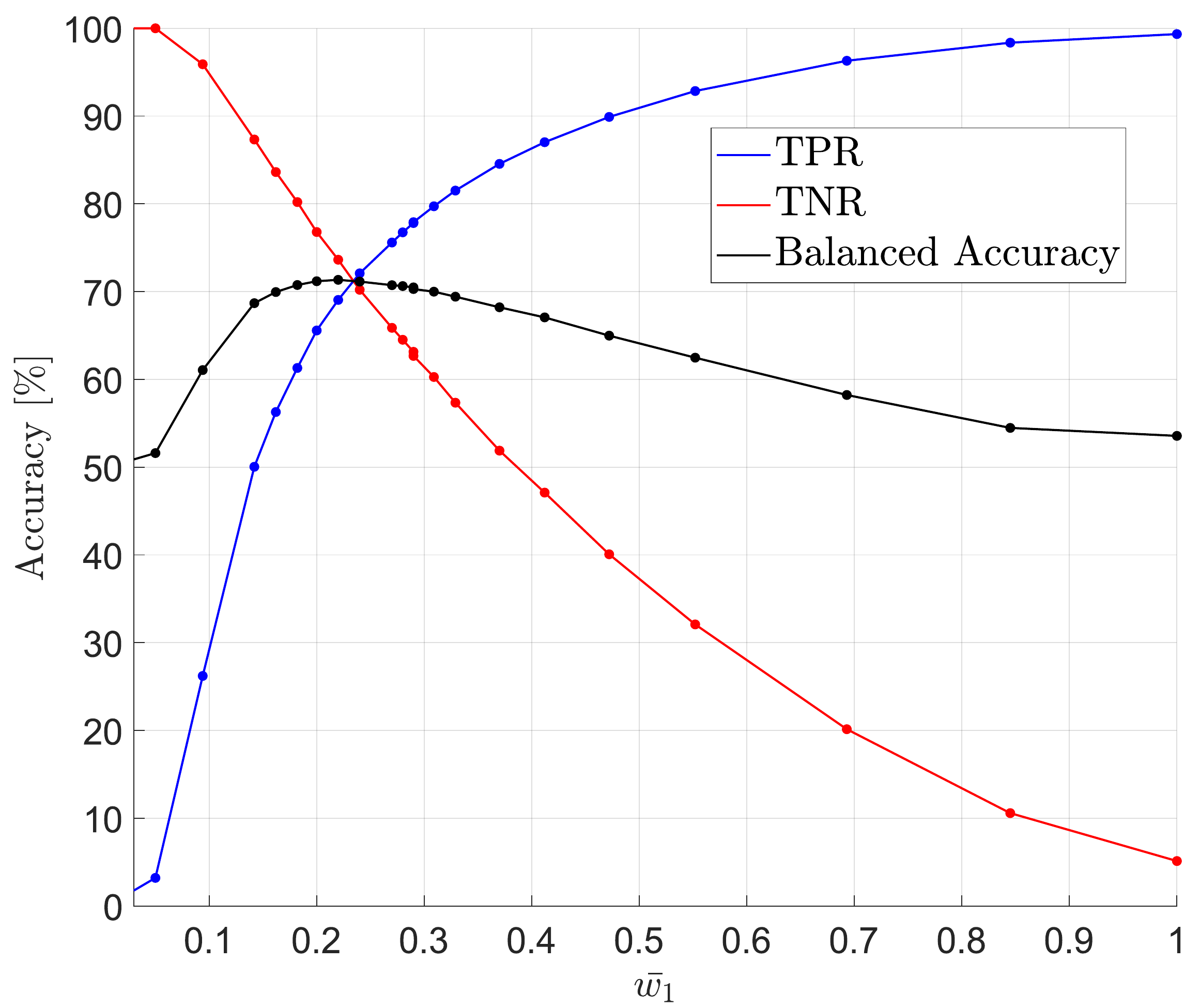}
			\caption{TNR, TPR, and balanced accuracy with respect to average class weight in the generalized scenario}
			\label{fig:omegablg}
		\end{figure}

\begin{figure}
			\centering
			\includegraphics[scale=0.35]{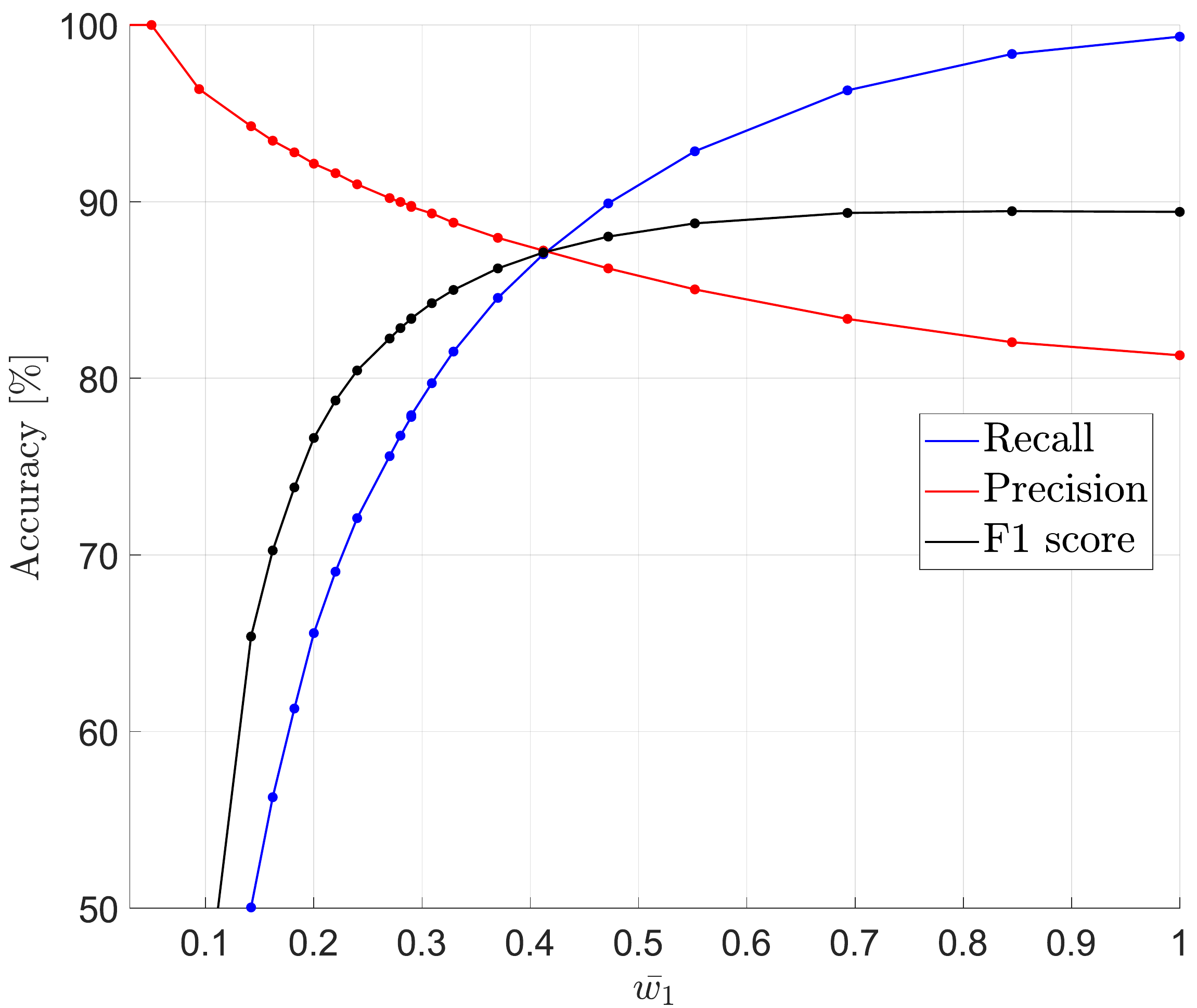}
			\caption{Precision, recall and F1 dynamics with respect to average class weight in the generalized scenario}
			\label{fig:omegaf1g}
		\end{figure}

		The neighborhood analysis and the performance variation with respect to kernel size $\sigma$ are depicted in Fig. \ref{fig:analg} and \ref{fig:vsg}. Performance dynamics regarding the variation of class weight $\omega_{1}$ are represented in Fig. \ref{fig:omegablg} and \ref{fig:omegaf1g}. These trends interestingly resemble those of the constrained case. So, the sensitivity of the prediction accuracy in terms of switching between the two scenarios is fairly robust. The complete trace of the performance metrics of the generalized case is reflected in Table \ref{tbl:2}. This table exhibits the effective application of our SVM-based algorithm to incorporate the notion of parity in the convergence prediction of the swarm.
		
\begin{figure}
			\centering
			\includegraphics[scale=0.35]{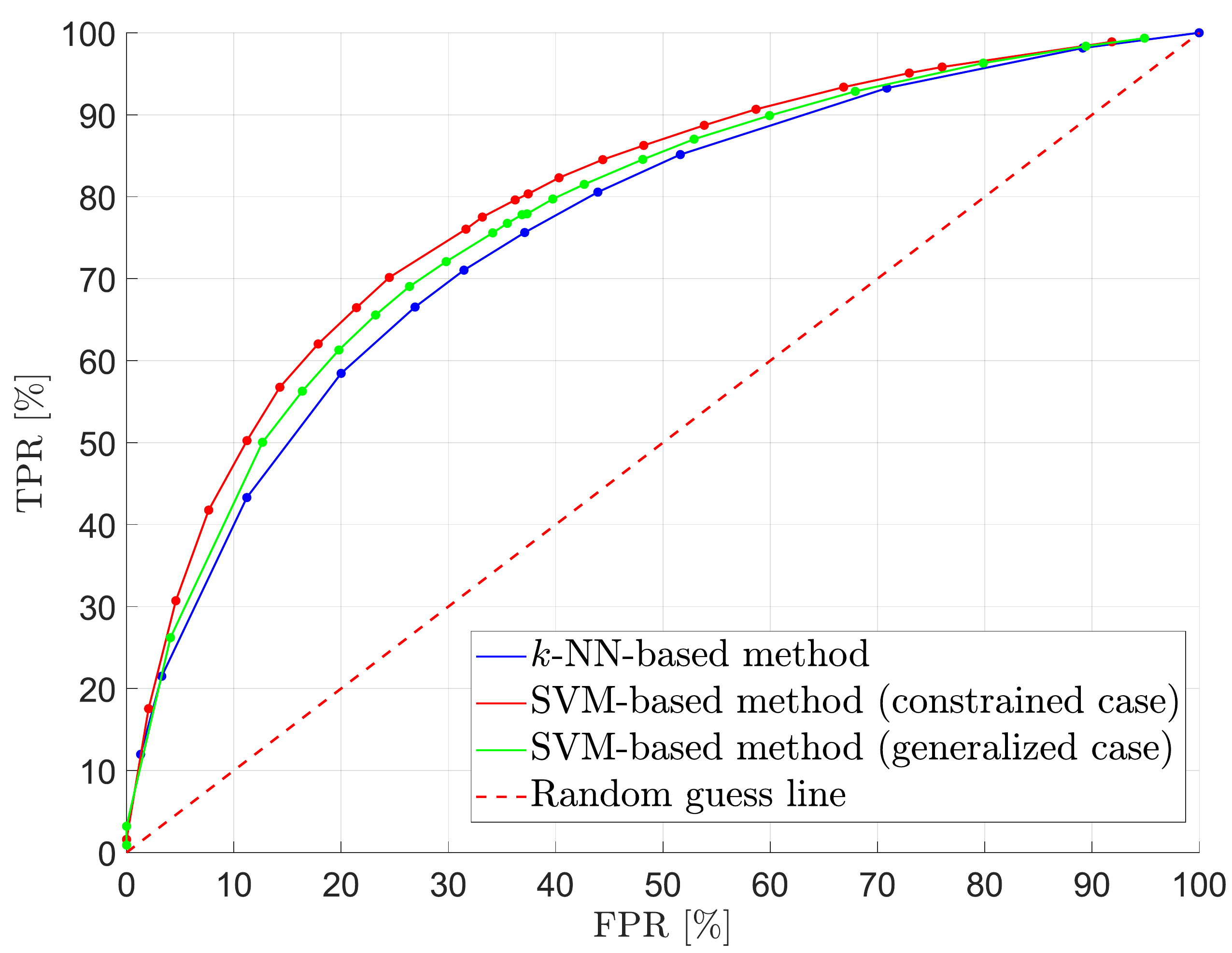}
			\caption{Overall comparative ROC curves of both the constrained and the generalized scenarios}
			\label{fig:rocg}
		\end{figure}

		The comparison of the predictive functionalities of all of the available algorithms, i.e., the constrained $k$-NN-based algorithm, the constrained SVM-based algorithm, and the generalized SVM-based algorithm, is illustrated in the ROC curve set of Fig. \ref{fig:rocg}. In particular, one observes that the constrained SVM-based approach is more expressive than the constrained $k$-NN-based method. In other words, with the assumption of fixed parities, the former has to be preferred to the latter. Moreover, the ROC curve of the generalized SVM-based strategy is above that of the constrained $k$-NN-based one, but trivially below that of the constrained SVM-based method. Comparatively, the generalized SVM-based algorithm deals with the complexity of parity which is not taken into account by the constrained version. Nevertheless, it is not an excessive cost in the performance reduction of the prediction by adding parity to predictions.
		\begin{table}
			\centering
			\caption{The best prediction results corresponding to the generalized case}
			\begin{tabular}{c@{\hskip 0.3in}c@{\hskip 0.3in}c@{\hskip 0.2in}c@{\hskip 0.2in}c@{\hskip 0.2in}c@{\hskip 0.2in}c@{\hskip 0.2in}c@{\hskip 0.2in}c}
				\toprule  
				Case&NT&$\omega_1$&$\sigma$&TPR(\%)&TNR(\%)&BA(\%)&Precision(\%)&F1(\%)\\\cmidrule{1-9}
				\multirow{5}{*}{I}&6&0.290&1.55&\multirow{5}{*}{75.1}&\multirow{5}{*}{66.6}&\multirow{5}{*}{70.8}&\multirow{5}{*}{90.3}&\multirow{5}{*}{82.0}\\
				&5&0.250&1.55&&&&&\\
				&4&0.200&1.17&&&&&\\
				&3&0.140&0.99&&&&&\\
				&2&0.110&0.99&&&&&\\\cmidrule{1-9}
				\multirow{5}{*}{II}&6&0.330&1.42&\multirow{5}{*}{79.7}&\multirow{5}{*}{60.5}&\multirow{5}{*}{70.1}&\multirow{5}{*}{89.4}&\multirow{5}{*}{84.3}\\
				&5&0.290&1.42&&&&&\\
				&4&0.240&1.05&&&&&\\
				&3&0.180&1.09&&&&&\\
				&2&0.150&1.09&&&&&\\
				\bottomrule		
			\end{tabular}
			\label{tbl:2}
		\end{table}
		\section{Conclusion}
		\label{sec:conc}
		Astrobot swarms are populated extremely dense formations of specific manipulators whose collision-free coordination are of utmost importance for astronomical operations. Cosmological operations often require that the number of fully coordinated astrobots is above a particular threshold. However, the convergence rates achieved by the distributed controllers of astrobots can only be studied using intensive simulations. The prediction of astrobots' convergences has been already done in a constrained case using a $k$-NN-driven strategy in which only spatial features of astrobots' targets are taken into account. In this paper, we illustrate that the accuracy performance of our SVM-based algorithm is higher than that of the $k$-NN-based one. Moreover, our algorithm also brings parity, say, the rotation direction of the outer arm of an astrobot, into play, thereby generalizing the solution to the convergence prediction problem. The comparative study of the performance results indicates that the parity addition to the formulation of our predictor only trivially reduces the quality of its prediction accuracy. So, one observes that the generalization of the convergence prediction is efficiently realized.
		
		An unexplored venue to potentially further improve our results would be the usage of convolutional neural networks (CNN). The existence of hidden layers may provide novel ideas to perceive more information about the intermediate coordination steps from an initial configuration to a final one by exclusively owning these two configurations. However, CNN-driven designs are often less intuitive than the designs based on more geometrical approaches such as $k$-NN and SVM algorithms. Specially, planning the number of  layers and the design of convolution computation and pooling operations are the challenges which have to be overcome.
		\subsection* {Acknowledgments}
		This work was financially supported by the Swiss National Science Foundation (SNF) Grant No. 20FL21\_185771 and the SLOAN ARC/EPFL Agreement No. SSP523. \z{The authors also appreciate the thoughtful comments and the suggestions of an anonymous reviewer and Dr. Ian N. Evans, i.e., the associate editor, on the earlier draft of this paper.} 
		
		
		\nocite{*}
		\bibliographystyle{IEEEtran}
		\bibliography{report}{}
\end{document}